\documentclass[twocolumn]{jpsj2}

\def\runtitle{Effects of degenerate orbitals on the Hubbard model
}
\def\runauthor{Akihisa \textsc{Koga}, Takuma \textsc{Ohashi}, 
Yoshiki \textsc{Imai}, Sei-ichiro \textsc{Suga} and Norio \textsc{Kawakami}}

\title{
Effects of degenerate orbitals on the Hubbard model
}

\author{%
Akihisa Koga, Takuma Ohashi, Yoshiki Imai, Sei-ichiro Suga and 
Norio Kawakami
}

\inst{
Department of Applied Physics, Osaka University, Suita, 
Osaka 565-0871, Japan
}

\recdate{\today}

\abst{%
}

\kword{%
degenerate Hubbard model, Coulomb interaction, quantum Monte Carlo simulation
}

\begin{document}
\sloppy
\maketitle

\begin{figure}[htb]
\begin{center}
\includegraphics[width=7cm]{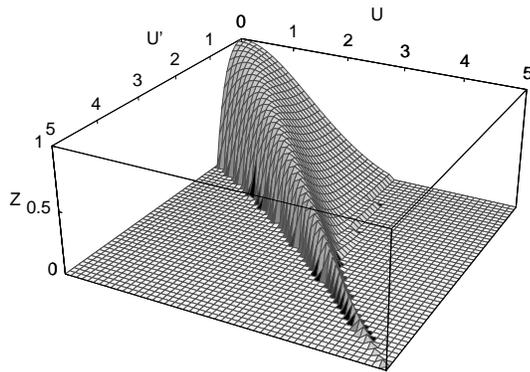}
\end{center}
\vskip -4mm
\caption{Quasi-particle weight for the degenerate Hubbard model 
calculated by the DMFT within the exact diagonalization.
$Z=0$ ($Z\neq 0$) corresponds to the 
insulating (metallic) phase.
}
\label{fig:phase}
\end{figure}

Strongly correlated electron systems with degenerate orbitals 
have attracted much interest. 
One of typical examples is the transition metal compound 
$\rm LiV_2O_4$,\cite{Kondo}
where the heavy-fermion behavior was observed 
down to very low temperatures. 
It was suggested that $t_{2g}$ orbitals as well as 
geometrical frustration are important to understand the ground state 
stabilized against other ordered states.\cite{Tsunetsugu}
Another example is the $f$-electron system $\rm CeTIn_5$ 
$\rm (T=Co, Ir, Rh)$,\cite{Hegger,Pagliuso,Takeuchi} 
where the superconducting state competes with the magnetically ordered state.
It was claimed that the effective Hamiltonian for this compound 
may be described by the degenerate Hubbard model.\cite{Takimoto}
Therefore, it is desirable to investigate low-energy properties 
of the degenerate Hubbard model systematically.
In our previous paper,\cite{Koga} 
we studied the ground-state properties 
of the degenerate Hubbard model 
by means of the dynamical mean field theory (DMFT)
combined with the exact diagonalization method.\cite{Georges}
It was found that if the intra- and inter-band Coulomb interactions 
are nearly equal, 
the metallic ground state is stabilized up to fairly large interactions 
due to orbital fluctuations.
Shown in  Fig. \ref{fig:phase}  is a more precise phase diagram 
newly obtained by DMFT with
the exact diagonalization. 
However, we were not able to systematically deal with dynamical properties 
in the previous paper because of the limitation inherent in
the exact diagonalization of small clusters.
We address this problem here by combining 
the quantum Monte Carlo simulations with DMFT. This also enables 
us to explore finite-temperature properties, which are useful to 
reveal the many-body character in correlated electron  systems.


We consider a correlated electron system 
with two-fold degenerate orbitals,
which may be described by the Hubbard Hamiltonian,
\begin{eqnarray}
H&=&\sum_{k,\alpha,\sigma}
\left(\epsilon_k-\mu\right) c_{k\alpha\sigma}^\dag c_{k\alpha\sigma}\nonumber\\
&+&U\sum_{i\alpha}n_{i\alpha\uparrow}n_{i\alpha\downarrow}
+U'\sum_{i,\sigma,\sigma'}n_{i1\sigma}n_{i2\sigma'}
\end{eqnarray}
where $c_{i\alpha\sigma}^\dag (c_{i\alpha\sigma})$ 
creates (annihilates) an electron 
with  spin $\sigma(=\uparrow, \downarrow)$ and orbital
$\alpha(=1, 2)$ at the $i$th site, 
and $n_{i\alpha\sigma}=c_{i\alpha\sigma}^\dag c_{i\alpha\sigma}$.
$\epsilon_k$ represents the dispersion relation, $\mu$ the chemical potential,
$U$ and $U'$ the intra-band and inter-band Coulomb interaction.
We use the semicircular density of states 
$\rho(x)=2/\pi D \sqrt{1-(x/D)^2}$ for simplicity.
We focus here on the symmetric case 
by setting  $\mu=\frac{1}{2}U+U'$.
In the following, we take the band width $D$ as unity.

To discuss how two kinds of Coulomb interactions affect the 
stability of the metallic state, 
we use DMFT,\cite{Georges} which can incorporate local electron correlations 
 precisely.
This method has been applied to the degenerate Hubbard model
by several groups.\cite{Momoi,QMC,Han,Koga,Ono}
In this paper, we exploit the quantum Monte Carlo simulations based on 
the Hirsch-Fye algorithm \cite{Hirsch}
to clarify finite-temperature properties 
of the degenerate model.

\begin{figure}[htb]
\begin{center}
\includegraphics[width=7cm]{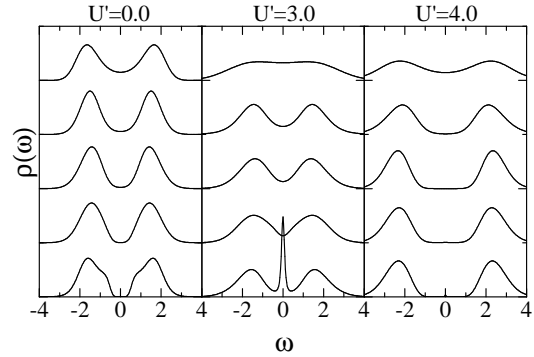}
\end{center}
\vskip -4mm
\caption{Density of states for the degenerate Hubbard model $(U=3.0)$. 
The data are for inverse 
temperature $T^{-1}=1, 2, 4, 8$ and $16$ from the top to the bottom.}
\label{fig:dos}
\end{figure}

In Fig. \ref{fig:dos}, we show the density of states deduced by 
applying the maximum entropy method to the Monte Carlo data.
When $U'=0.0$, the system is reduced to the single-band Hubbard model, 
where the metal-insulator transition occurs at a critical point 
$U_c=2.94$.\cite{Bulla}
In the case  $(U, U')=(3.0, 0.0)$, the system is in the insulating phase 
close to the transition point. 
Nevertheless we can clearly observe the formation of the Hubbard gap
with the decrease of temperature $T$.
Introducing the interband Coulomb interaction $U'$,
the sharp quasi-particle peak appears around 
the Fermi surface in the case $(U, U')=(3.0, 3.0)$.
This implies that orbital fluctuations induced by 
the interband Coulomb interaction drives the 
system to the metallic phase.
\begin{figure}[htb]
\begin{center}
\includegraphics[width=6cm]{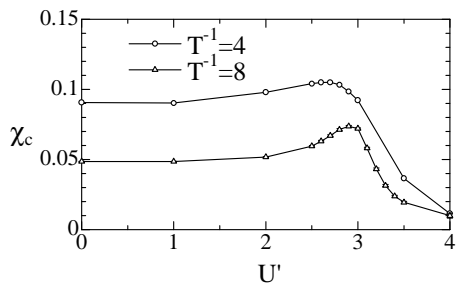}
\end{center}
\vskip -4mm
\caption{Local charge susceptibility as a function of 
the interband Coulomb interaction with a fixed $U=3.0$.}
\label{fig:chi-c}
\end{figure}
In fact, the local charge susceptibility is increased up to $U\sim U'$ 
at low temperatures, as seen in Fig. \ref{fig:chi-c}.
These results are consistent with the previous results.\cite{Han,Koga,Ono}
Further increase in the interband interaction suppresses  
 spin fluctuations, leading 
the system to another type of the Mott insulator in the 
region of $U'>U$ (see Fig. \ref{fig:phase}).

\begin{figure}[htb]
\begin{center}
\includegraphics[width=8cm]{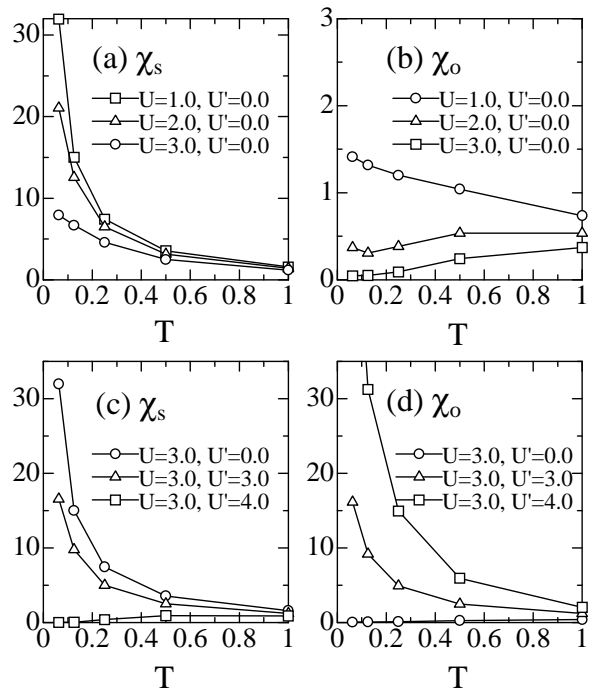}
\end{center}
\vskip -4mm
\caption{Local spin and orbital susceptibilities 
as a function of temperature.}
\label{fig:chi}
\end{figure}

To characterize the nature of two types of the Mott insulators,
we investigate the temperature dependence of the 
local spin and orbital susceptibilities,
which are defined as,
\begin{eqnarray}
\chi_s&=&\int_0^\beta {\rm d}\tau\langle
\left\{n_\uparrow(0)-n_\downarrow(0)\right\}
\left\{n_\uparrow(\tau)-n_\downarrow(\tau)\right\}\rangle\nonumber\\
\chi_o&=&\int_0^\beta {\rm d}\tau\langle
\left\{n_1(0)-n_2(0)\right\}
\left\{n_1(\tau)-n_2(\tau)\right\}\rangle,
\end{eqnarray}  						  
where $\beta=T^{-1}$, $n_\sigma=\sum_\alpha n_{\alpha,\sigma}$, 
$n_\alpha=\sum_\sigma n_{\alpha, \sigma}$ and $\tau$ is an imaginary time.
We show the results obtained by the quantum Monte Carlo simulations
within DMFT in Fig. \ref{fig:chi}. 
Let us first look at Figs. \ref{fig:chi} (a) and (b) for  $U'=0$ 
(equivalent to the single band model). Since
the introduction of the intra-orbital interaction $U$  sharpens 
the quasi-particle peak, this gives rise to the increase of 
the spin susceptibility $\chi_s$ at low temperatures. On 
the other hand, the formation of the Hubbard gap 
suppresses not only the charge susceptibility,
but also the orbital susceptibility $\chi_o$.  
As seen from Figs. \ref{fig:chi} (c) and (d),
we encounter quite different behavior in the susceptibilities, 
when   the interband interaction $U'$ is increased. 
Namely, the spin susceptibility is suppressed, while
the orbital susceptibility is enhanced
at low temperatures. 
We have checked that this tendency  holds 
 for larger $U'$ beyond the condition $U' \sim U$.

Therefore in the metallic phase close to the Mott insulator
in the region of  $U>U' (U<U')$, spin (orbital) fluctuations 
are enhanced  whereas orbital (spin) fluctuations are suppressed with 
the decrease of temperature.
These analyses shed some light on the reason 
why the metallic phase is particularly stable along the 
line $U=U'$.  On this line, spin and orbital fluctuations
are almost equally enhanced, and this subtle balance is efficient
to stabilize the metallic phase. 
When the ratio of interactions deviates from this condition, the 
system prefers either of two types of the Mott insulating phases.

In this paper, we have restricted our 
discussions to the paramagnetic phase.
As mentioned above, spin and/or orbital fluctuations are
particularly enhanced
in the metallic phase close to the Mott insulator, which 
would trigger a quantum phase transition to a certain 
ordered phase.\cite{Momoi}  This problem is to be addressed in the 
future work. Nevertheless, we think that the 
general feature about the stability of 
metallic phase  near $U \sim U'$ may hold even 
in those cases.




The authors thank the Yukawa Institute for Theoretical Physics at 
Kyoto University. 
Discussions during the YITP workshop YITP-W-02-17 on 
"Development of the Theory of Strongly Correlated Electrons with 
Orbital Degrees of Freedom" were useful to complete this work.
This work was partly supported by a Grant-in-Aid from the Ministry 
of Education, Science, Sports and Culture of Japan. 
A part of computations was done at the Supercomputer Center 
at the Institute for Solid State Physics, University of Tokyo
and Yukawa Institute Computer Facility.

\end{document}